\documentclass[%
aip,
jcp,
 amsmath,amssymb,
reprint,%
]{revtex4-2}

\usepackage{graphicx}% Include figure files
\usepackage{dcolumn}% Align table columns on decimal point
\usepackage{bm}% bold math
\usepackage{xcolor}			% Color some parts red to focus attention
\usepackage{braket}			% Braket notation

\usepackage[utf8]{inputenc}
\usepackage[T1]{fontenc}
\usepackage{mathptmx}
\usepackage{etoolbox}

\makeatletter
\def\@email#1#2{%
 \endgroup
 \patchcmd{\titleblock@produce}
  {\frontmatter@RRAPformat}
  {\frontmatter@RRAPformat{\produce@RRAP{*#1\href{mailto:#2}{#2}}}\frontmatter@RRAPformat}
  {}{}
}%
\makeatother
\begin{document}

\preprint{AIP/123-QED}

\title{Optical Signatures of the Coupling between Excitons and Charge Transfer States in Linear Molecular Aggregates}
% Force line breaks with \\
\author{M. Manrho}
\affiliation{%
	University of Groningen, Zernike Institute for Advanced Materials, Nijenborgh 4, 9747 AG Groningen, The Netherlands%\\This line break forced% with \\
}%
\author{T. L. C. Jansen}%
\affiliation{%
	University of Groningen, Zernike Institute for Advanced Materials, Nijenborgh 4, 9747 AG Groningen, The Netherlands%\\This line break forced% with \\
}%
\author{J. Knoester}%
 \email{j.knoester@rug.nl.}
 \homepage{https://www.rug.nl/staff/j.knoester/}
\affiliation{%
University of Groningen, Zernike Institute for Advanced Materials, Nijenborgh 4, 9747 AG Groningen, The Netherlands%\\This line break forced% with \\
}%
\affiliation{Leiden University, Faculty of Science, Einsteinweg 55, 2300 RA Leiden, The Netherlands}

\date{\today}% It is always \today, today,

\begin{abstract}
Charge Transfer (CT) has enjoyed continuous interest due to increasing experimental control over molecular structure leading to applications in, for example, photovoltaics and hydrogen production. In this paper, we investigate the effect of CT states on the absorption spectrum of linear molecular aggregates using a scattering matrix technique that allows us to deal with arbitrarily large systems. The presented theory performs well for both strong and weak mixing of exciton and CT states, bridging the gap between previously employed methods which are applicable in only one of these limits. In experimental spectra the homogeneous linewidth is often too large to resolve all optically allowed transitions individually, resulting in a characteristic two-peak absorption spectrum in both the weak- and strong-coupling regime. Using the scattering matrix technique we examine the contributions of free and bound states in detail. We conclude that the skewness of the high-frequency peak may be used as a new way to identify the exciton-CT-state coupling strength.
\end{abstract}

\maketitle

%%%
%%% Introduction
%%%
\section{\label{sec:introduction}Introduction}
%%%
%%% Catchy introduction
%%%
Crystal structures of organic molecular aggregates have become more refined over the last decade. By clever choice of side chains the molecular packing can be imposed with great precision, which allows one to study the effects of tight molecular packing on functional properties, such as optical absorption and excitation energy transport. Already in the previous century the effects of close packing have been linked to the formation of charge transfer (CT) states. \cite{merrifield_ionized_1961, samoc_photoconductivity_1983, hennessy_vibronic_1999} With CT states being found recently in new materials\cite{ herbst_self-assembly_2018, gao_vibronic_2011, weingarten_self-assembling_2014, rinn_interfacial_2017, scholz_time-resolved_2004} and their appearance in biological light-harvesting systems,\cite{novoderezhkinMixingExcitonChargeTransfer2007, cupelliniCouplingChargeTransfer2018a, cupelliniChargeTransferCarotenoid2020, maDynamicStarkEffect2021} interest in modeling of CT systems persisted \cite{gisslen_crystallochromy_2009, liExcitedStateStructureModifications2014, petelenz_charge-transfer_2015, kubo_parameter-free_2018,  oleson_perylene_2019} which fuelled the need for a better understanding how the competition between various interactions leads to specific observable properties.

%%%
%%% The goal of the paper
%%%
This paper focuses on the effects of CT states on the absorption spectrum of linear molecular aggregates. Using a scattering matrix technique we obtain a method applicable for both strong and weak mixing of Frenkel and CT exciton states. This method enables the identification of bound and free states in the absorption spectrum. We show that, in principle, this method allows one to obtain the mixing strength between Frenkel and CT exciton states from the absorption spectrum alone, without having to perform quantum chemical calculations. Furthermore, by applying the presented method to experimental spectra and comparing to results obtained using traditional methods of direct Hamiltonian diagonalization, we conclude that the latter can lead to ambiguous interpretations.

%%%
%%% General introduction
%%%
Organic molecular aggregates are often formed through van der Waals interactions, in contrast to the covalent bonds found in inorganic compounds. As a result, the molecular orbitals (MOs) of the individual molecules are a good approximation of the electronic states of the aggregate. For this reason, optical (charge neutral) excitations in these materials often can be understood as linear combinations of single-molecule excited states, i.e., states in which a single molecular excitation quantum is shared coherently by many molecules. Such states are known as Frenkel excitons; their delocalization is driven by resonance excitation transfer interactions, usually of dipolar origin. This coupling leads to the well known H- and J-aggregate behavior as described by the Kasha model. \cite{kasha_energy_1963}

During the last decades, several tightly packed molecular structures have been obtained in experiments. \cite{herbst_self-assembly_2018, gao_vibronic_2011, weingarten_self-assembling_2014, rinn_interfacial_2017, scholz_time-resolved_2004} Tight packing of molecules, e.g., with inter-molecular distances of $\sim 3$\AA,  causes the MOs of neighboring molecules to overlap. \cite{hestand_exciton_2015} This overlap gives rise to additional couplings, referred to as CT couplings, where an electron is transferred from an excited molecule to its neighbor, leaving a negatively and a positively charged molecule. In solid-state physics language, this may be regarded as an electron and a hole, which subsequently may move further away from each other through CT couplings with other molecules. The Coulomb attraction between the electron and hole, the third relevant interaction in these type of systems, creates bound states which are reflected as individual peaks in the absorption spectrum. Many characteristics of Wannier excitons in traditional inorganic semiconductors also apply to CT states. For example, the spectrum of bound state solutions of CT systems bears close resemblance to the hydrogen spectrum. \cite{bounds_charge-transfer_1980, forrest_ultrathin_1997}

The combination of the three interactions introduced above, i.e., resonant excitation transfer interactions, CT couplings, and Coulomb attraction, results in an intricate Hamiltonian that is hard to solve exactly. In one dimension this Hamiltonian can be solved using hypergeometric functions, \cite{merrifield_ionized_1961} but this method does not lead to workable expressions for practical applications. Hence, in the last couple of decades one often diagonalized the Hamiltonian using numerical techniques, \cite{so_evidence_1991, hennessy_vibronic_1999, hestand_interference_2015, hestand_extended-charge-transfer_2016} which even in absence of vibrations is numerically challenging, because the Hamiltonian dimension grows quadratically with the system size. In order to keep the calculation computationally feasible, the basis set has to be truncated by limiting the number of molecules and by restricting the maximum charge separation of the CT states. For some materials, the charge separation can even be restricted to nearest neighbors, \cite{hoffmann_lowest_2000, lalov_vibronic_2007, eilmes_charge_1991, oleson_perylene_2019, hoffmann_optical_2002} while retaining good agreement with experimental results. However, restricting the maximum charge separation effectively deepens the Coulomb well depth, thus complicating the interpretation of model parameters. In this paper we avoid restriction of the charge separation by truncating the Coulomb interaction instead.

Here we use a Green's function method \cite{economou_greens_1983} to calculate the absorption spectrum, where the Coulomb attraction between the electron and hole is treated using a scattering matrix technique. With this method there is no need to truncate the basis, which allows us to study an infinitely long one-dimensional chain. The scattering matrix technique enables us to discern bound states from continuum states by investigating the poles of this matrix.\cite{liExcitedStateStructureModifications2014} We use this distinction to improve our understanding of the resulting absorption spectrum.

The absorption spectrum is dominated by Frenkel excitons, as they posses the largest transition dipole moment and hence, the effect of CT states on the absorption spectrum depends on the mixing between the Frenkel- and CT excitions. This mixing is determined by the ratio of the CT coupling and the Coulomb well depth. We will show that, in the limit of a large homogenous linewidth, the weak, intermediate, and strong mixing regimes all have a corresponding two-peak absorption spectrum. This result is in line with previous studies which found a two-peak spectrum in a dimer \cite{hennessy_vibronic_1999} as well as for larger aggregates. \cite{oleson_perylene_2019, hestand_interference_2015, lalov_vibronic_2007, petelenz_charge-transfer_2015} We find that one of the peaks may always be ascribed to a bound state, while the second peak can result from a single bound state, from continuum states, or from a combination of the two. We find a direct relationship between the exciton-CT-state mixing and the energy position and lineshape of the second peak, providing a new tool to estimate the exciton-CT-state coupling strength from absorption spectra.

%%%
%%% Outline
%%%
The structure of this paper is as follows. In Section~\ref{sec:model} we first introduce the model Hamiltonian used to describe the system. Then we derive the unperturbed Green function and the scattering matrix arising from the Coulomb interaction. The Green function is solved in Section~\ref{sec:results} where we consequently explore the parameter space of the model and show how we can use the second peak in the absorption spectrum as a proxy for exciton-CT-state mixing. Finally, we conclude in Section~\ref{sec:Conclusion}. Several details of derivations and the computation are described in appendix A and B.

%%%
%%% Model and Hamiltonian
%%%
\section{\label{sec:model} Model and Method}
\subsection{\label{sec:hamiltonian} Model Hamiltonian}
While the scattering method presented below may in principle be applied equally well to higher-dimensional systems, for simplicity and ease of interpretation we consider a one-dimensional model system. Here, it is relevant to note that it has been shown that the one-dimensional chain provides a good approximation for many organic crystals exhibiting CT couplings. \cite{hestand_expanded_2018, gisslen_crystallochromy_2009, hoffmann_lowest_2000, gao_vibronic_2011, yamagata_hj-aggregate_2014} We consider the chain to be infinitely long and consisting of identical molecules. The system thus exhibits translational invariance and contains one molecule per unit cell. 

Our focus lies on the  linear absorption spectrum and we assume the relevant molecular electronic orbitals are the HOMO and the LUMO and only consider states with a single electronic excitation. The molecular ground state is the fully occupied HOMO. A dipole allowed electronic excitation on a single molecule brings an electron from the HOMO to the LUMO, leaving two partially filled orbitals. In solid state physics terminology, a partially filled LUMO and HOMO may be considered to represent an electron and hole, respectively. A general basis for this system is spanned by $\ket{n;s}$, where $n = 0,1,2, \cdots$ denotes the position of the electron and $s = 0, \pm1, \pm 2,\cdots$ is the position of the hole relative to the electron. When $s=0$, the state describes a molecular (local) excitation and when $s\neq 0$, it describes a CT state. \cite{lang_kinetic_1963}

Conventionally, the Hamiltonian is constructed by extending the exciton part of the Hamiltonian to include CT states.\cite{hestand_expanded_2018} Here we take a different approach and write the total Hamiltonian as $H = H_0 + H'$, where the unperturbed Hamiltonian $H_0$ considers the electron and hole to be completely free and independent particles governed by the CT couplings and $H' $ contains all the remaining interactions. Explicitly we have:
\begin{equation}
    \begin{split}
        H_0 = & \sum_{n,s} E_{\text{CT}}\ket{n;s}\bra{n;s} \\
        &+ \sum_{n,s,\pm} \left( t_{\text{e}} \ket{n;s}\bra{n\pm 1; s \mp 1} + t_{\text{h}} \ket{n;s}\bra{n;s\pm 1}\right), 
    \end{split}
\end{equation}
where the energy of a state in the site basis is denoted $E_{\text{CT}}$, which in the real system, including Coulomb interactions, represents the energy of a state with infinite charge separation. Due to wave function overlap between neighboring molecules, the electron and hole can move between sites. These CT couplings, denoted with $t_{\text{e}}$ and $t_{\text{h}}$ for the electron and hole respectively, are defined as the matrix elements of the full electronic Hamiltonian $H_{\text{el}}$ between the LUMO or HOMO orbitals on neighboring sites, i.e.,
\begin{equation}
    t_{\text{e}} = \bra{LUMO_1}H_{\text{el}}\ket{LUMO_2},
\end{equation}
\begin{equation}
    t_{\text{h}} = \bra{HOMO_1}H_{\text{el}}\ket{HOMO_2}.
\end{equation}
The CT couplings only extend to nearest neighbors and are highly sensitive to the molecular packing.\cite{kazmaier_theoretical_1994, hestand_exciton_2015}

\begin{figure}
	\centering
	\includegraphics[width=8.5cm]{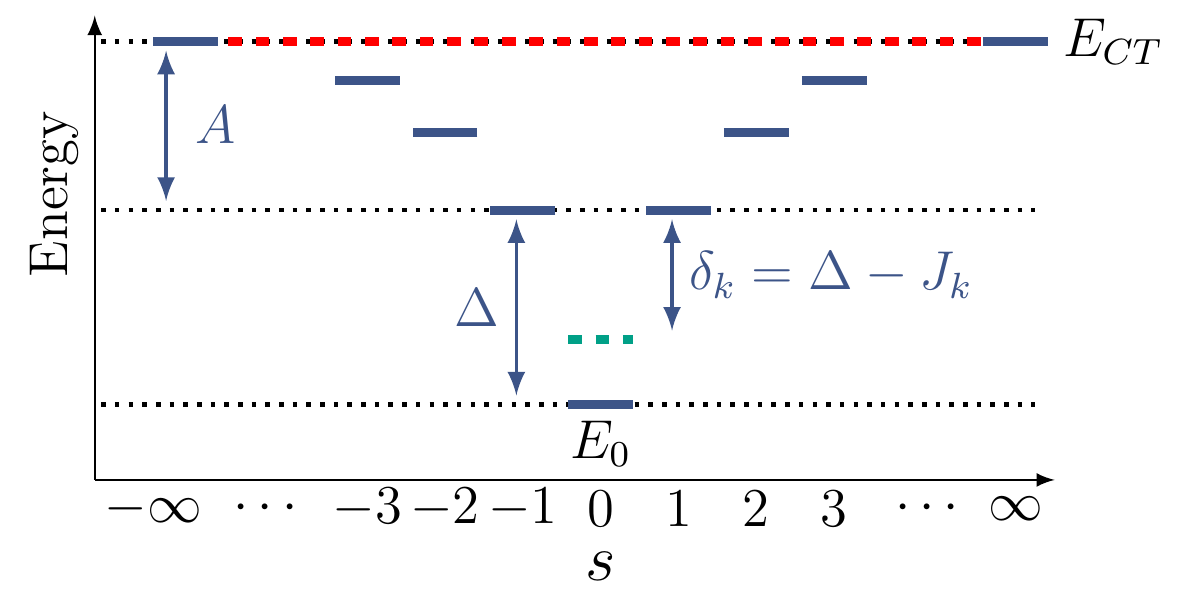}
	\caption{Energy level diagram indicating the energies of the basis states $\ket{n;s}$. The dashed red line denotes the energy $E_{\text{CT}}$ of states with infinite charge separation. The molecular excited state, corresponding to zero charge separation $s$, generally has the lowest energy ($E_0$). The nearest-neighbor CT states ($|s|=1$) are detuned from the molecular excitation energy by $\Delta$ and the Coulomb well depth is indicated by $A$. As a consequence of the resonant excitation transfer interaction $J_{n,m}$ (see Eq.~(\ref{eq:perturbationHamiltonian})), the single-molecule excitation energy is shifted by an amount $J_k$  (see Eq.~(\ref{eq:fourierDipole}) below; $k$ the total quasi-momentum) to the exciton energy (green dashed line), which alters the detuning to $\delta_k$. }
	\label{fig:EnergyLevelDiagram}
\end{figure}

The perturbation Hamiltonian $H' $ contains the Coulomb attraction $V(s)$ of the electron-hole pair and the resonant excitation transfer interactions $J_{n,m}$ between excitations on molecules $n$ and $m$,
\begin{equation}
H' = \sum_{n,s} V(s)\ket{n;s}\bra{n;s} + \sum_{n \neq m} J_{n,m} \ket{n;0}\bra{m;0}.
\label{eq:perturbationHamiltonian}
\end{equation}
The Coulomb potential is given by
\begin{equation}
V(s) = -\left(A+\Delta \right)\delta_{s,0} - \frac{A}{s} \left(1-\delta_{s,0}\right),
\label{eq:Coulombwell}
\end{equation} 
where $\Delta$ is the energetic detuning between a molecular excitation and a nearest-neighbor CT state and $A$ denotes the Coulomb well depth. Thus, the excitation energy of a single molecule may be written $E_0 = E_{\text{CT}}-A-\Delta$. This energy level and the Coulomb potential are depicted in Fig.~\ref{fig:EnergyLevelDiagram} by the solid blue lines. On length scales of several \AA\ a screened Coulomb potential as in Eq.~(\ref{eq:Coulombwell}) provides a fairly good approximation for the interactions between orbitals of neutral molecules.\cite{forrest_ultrathin_1997, yamagata_hj-aggregate_2014, hernandez_optical_1969, bounds_charge-transfer_1980, petelenzDipoleMomentValid2013} The interactions $J_{n,m}$ usually have a large dipolar contribution, but they may also contain contributions from higher-order multipoles. For the purpose of this paper, no specific origin needs to be specified.

As will be described in the next sections, the total Hamiltonian $H$ allows us to construct the Green function and consequently find the absorption spectrum. 

%%%
%%% Absorption spectrum
%%%
\subsection{Absorption spectrum}
The absorption spectrum $I(\omega)$ may be obtained from the auto-correlation function of the dipole operator,\cite{kubo_statistical-mechanical_1957, mukamel_principles_1999, didraga_optical_2004} which leads to
\begin{equation}
I(\omega) = \text{Im}\left(\bra{g}\hat{\mu} G(\omega)\hat{\mu}\ket{g}\right).
\label{eq:absorption}
\end{equation}
Here, $\omega$ denotes the energy of the incoming photon ($\hbar = 1$), $\ket{g}$ is the chain's ground state (in which all molecules have a doubly occupied HOMO), $G(\omega)$ is the Green function (see below), and $\hat{\mu}$ denotes the dipole operator of the system
\begin{equation}
\hat{\mu} = \mu \sum_n \left(\ket{g}\bra{n;0} + \ket{n;0}\bra{g}\right),
\end{equation}
where $\mu$ is the transition dipole of a single molecule (all equal in magnitude and orientation). The above dipole operator only accounts for the transition dipole between the molecular ground and excited states. The wave function overlap between molecules may result in nearest-neighbor CT states to also obtain a non-zero transition dipole moment from the ground state. \cite{hennessy_vibronic_1999, hoffmann_lowest_2000, lalov_model_2008} This CT transition dipole moment, however, is often an order of magnitude smaller than that of excitons.\cite{scholzResonantRamanSpectra2011} Therefore, we will neglect this contribution, as is commonly done. \cite{hestand_exciton_2015, hennessy_vibronic_1999} Within this approximation, CT states have to mix with exciton states to obtain oscillator strength and the absorption spectrum directly probes the mixing between CT states and excitons.

The Green function in Eq.~(\ref{eq:absorption}) is the one-sided Fourier transform of the time evolution operator and is given by 
\begin{equation}
G(\omega) = (H - \omega - i\Gamma)^{-1},
\label{eq:gf}
\end{equation}
where $H$ is the total Hamiltonian. \cite{economou_greens_1983} The term $-i\Gamma$ describes dephasing originating from system-bath interactions and induces a Lorentzian absorption lineshape with a full width at half maximum (FWHM) of $2\Gamma$ for every isolated eigenstate of $H$.

%%%
%%% Green function
%%%
\subsection{\label{sec:GreenFunction} Green function}	
In this subsection, we derive an expression for the Green function. Using $H = H_0 + H' $ we may write \cite{economou_greens_1983}
\begin{equation}
G = G_0 + G_0H'G_0 + G_0H'G_0H'G_0 + \cdots = G_0 + G_0 T G_0,
\label{eq:Greenfunction}
\end{equation}
where $G_0 = (H_0 - \omega - i\Gamma)^{-1}$ describes the propagation of a free electron and hole and the t-matrix $T$ describes the scattering on the Coulomb interactions $H'$ (Eq.~(\ref{eq:perturbationHamiltonian})).

The unperturbed Green function $G_0(\omega)$ can be obtained analytically. As we are interested in the limit of long chains, we may employ periodic boundary conditions to diagonalize $H_0$. This results in performing two Fourier transforms and a relocation of the center of mass of the electron and the hole.\cite{merrifield_ionized_1961} The first Fourier transform is given by
\begin{equation}
\ket{k;s}' = \frac{1}{\sqrt{N}} \sum_n e^{ ik\left(\frac{2n+s}{2}\right)} \ket{n;s}, \quad k = 0, \pm \frac{2\pi}{N},\pm \frac{4\pi}{N}, \cdots, \pi ,
\label{eq:Fourrier}
\end{equation}
where $N$ is the number of molecules in the chain. For odd $N$ the $k=\pi$ state is omitted. Note that we Fourier transform with respect to the center of mass coordinate $(n+s/2)$ of the electron and hole. Thus, $k$ denotes the wavenumber of the center of mass. The matrix elements of $H_0$ are then further simplified by relocating the center of mass with the phase factor $\alpha_k = \tan^{-1}\left(\frac{t_{\text{e}}-t_{\text{h}}}{t_{\text{e}}+t_{\text{h}}}\right)$ as follows \cite{merrifield_ionized_1961} 
\begin{equation}
\ket{k;s} = e^{i\alpha_k s}\ket{k;s}'.
\label{eq:comtrans}
\end{equation}

After the above two transformations, $H_0$ is block diagonal in $k$ and its non-zero matrix elements are:
\begin{equation}
\bra{k;s}H_0\ket{k;s\pm 1} = \epsilon_k,
\end{equation}
\begin{equation}
\bra{k;\pm\left(N/2+1\right)}H_0\ket{k;\mp N/2}  = \epsilon_k e^{\pi i k},
\label{eq:boundary}
\end{equation}
\begin{equation}
\bra{k;s} H_0 \ket{k;s} = E_0 + E_{\text{CT}},
\end{equation}
where $\epsilon_k = \sqrt{t_{\text{e}}^2 + t_{\text{h}}^2 + 2t_{\text{e}} t_{\text{h}}\cos(2\pi k/N)}$. 
Further diagonalization is achieved through a second Fourier transform with respect to the charge separation $s$. Here, special care is to be taken with the $k$ dependent periodic boundary condition of Eq.~(\ref{eq:boundary}). The eigenstates of $H_0$ are then found to be\cite{merrifield_ionized_1961} 
\begin{equation}
\ket{k;\nu} = \frac{1}{\sqrt{N}} \sum_s e^{i\nu s} \ket{k;s}  ,
\label{eq:evenfunctions}
\end{equation}
where for even values of $k$ the wavenumbers are $\nu = 0, \pm \frac{2\pi}{N}, \pm \frac{4\pi}{N}, \cdots, \pi$ and for odd values of $k$ we have a M\"obius like Hamiltonian\cite{heilbronner_huckel_1964} and obtain $\nu = \pm \frac{\pi}{N}, \pm \frac{3\pi}{N}, \cdots, \pm \frac{N-1}{N}\pi$. Again, for odd $N$ the wavenumbers $\nu = \pi$ and $\nu = \frac{N-1}{N}\pi$ are omitted, respectively. The corresponding eigenenergies are
\begin{equation}
\lambda(k;\nu) = E_{\text{CT}} + 2\epsilon_k \cos(\nu).
\label{eq:oddenergies}
\end{equation}

Using the eigen-decomposition of $H_0$ the unperturbed Green function may now be written as
\begin{equation}
G_0(\omega) = \sum_{k,\nu} \frac{\ket{k;\nu}\bra{k;\nu}}{\lambda(k,\nu) - \omega - i\Gamma}.
\label{eq:GzeroSum}
\end{equation}
Taking the limit of $N\rightarrow\infty$ transforms the summations over $k$ and $\nu$ into integrations. The poles of $G_0(\omega)$ now form a branch cut on the real axis. The band of $H_0$ thus becomes continuous, spanning the interval $[-2\epsilon_k, +2\epsilon_k]$. For both even and odd $N$ the matrix elements of $G_0$ are given by
\begin{equation}
\bra{k;s}G_0(\omega)\ket{k';r} = \frac{1}{2\pi} \int_{-\pi}^{\pi} d\nu \frac{e^{i\nu (s-r)}}{\omega + i\Gamma - E_{\text{CT}} - 2\epsilon_k\cos(\nu)}\delta_{k,k'}
\end{equation}
The integral has two solutions depending on $\omega$ lying inside or outside the band of $H_0$.\cite{economou_greens_1983} For $\omega$ outside the band, i.e., $|x| \equiv |\left(\omega-E_{\text{CT}}\right)/2\epsilon_k| > 1$,
\begin{equation}
\bra{k;s}G_0(\omega)\ket{k';r} = \frac{\delta_{k,k'}}{\sqrt{(\omega-E_{\text{CT}})^2-4\epsilon_k^2}}\left(x-\sqrt{x^2-1}\right)^{|s-r|}
\label{eq:Gzero_outside}
\end{equation}
while for $\omega$ inside the band ($|x|<1$) we have
\begin{equation}
\bra{k;s}G_0(\omega)\ket{k';r} = \frac{- i\delta_{k,k'}}{\sqrt{4\epsilon_k^2-(\omega-E_{\text{CT}})^2}}\left(x-\sqrt{1-x^2}\right)^{|s-r|}.
\label{eq:Gzero}
\end{equation}

In order to obtain the full Green function $G(\omega)$, we need the $T$ matrix in Eq.~(\ref{eq:Greenfunction}), which is defined as
\begin{equation}
\begin{split}
T(\omega) &= H' + H'G_0(\omega)H' + H'G_0(\omega) H'G_0(\omega) H' + \cdots \\
&= H' \left(1-G_0(\omega) H' \right)^{-1}
\end{split}
\label{eq:ScatteringMatrix}
\end{equation}
To find the $T$ matrix, we first note that $H' $ is diagonal on the basis $\ket{k,s}$ defined by Eq.~(\ref{eq:comtrans})
\begin{equation}
\bra{k;s}H'\ket{p;r} =  \delta_{k,p}  \delta_{s,r} \left[\left(E_{\text{CT}} + V(s)\right) + \delta_{s,0} J_{k} \right],
\end{equation}
where $J_k$ is the Fourier-transformed resonant excitation transfer interaction $J_{n,m}$, 
\begin{equation}
J_{k} = \sum_{n\neq 0} e^{ikn} J_{0,n},
\label{eq:fourierDipole}
\end{equation}
where we used the translational invariance.
Hence, a single Fourier transform turns the excitation transfer interactions $J_{n,m}$ into an energy term $J_k$ for the exciton states, as is well-known for Frenkel exciton systems.\cite{davydovTHEORYMOLECULAREXCITONS1964} The energy difference between the exciton and nearest-neighbor CT state after the Fourier transformation is a relevant quantity, which hereafter we refer to as the detuning $\delta_k = \Delta - J_k$. The energy of the exciton state including the effect of $J_k$ is depicted in Fig.~\ref{fig:EnergyLevelDiagram} by the dashed green line.

From Eq.~(\ref{eq:ScatteringMatrix}), we observe that if we consider all interactions in $H'$, the calculation of $T(\omega)$ requires inverting an $N\times N$ matrix for every $k$, which means that the introduction of the perturbation $H'$ and the $T$ matrix does not really simplify the evaluation of $G(\omega)$ compared to direct diagonalization of the full Hamiltonian for every $k$. However, it is possible to take advantage of the separation of $H_0$ and $H' $ by truncating the range of $H'$ and only considering contributions to $H'$ with $|s| \leq s_{\text{trun}}$. This boils down to ignoring the Coulomb attraction between the electron and hole for distances larger than $s_{\text{trun}}$. As we will show below, this allows us to obtain $G(\omega)$ using the inversion of matrices of dimension $2s_{\text{trun}} + 1$, independent of system size. 

The truncated Coulomb potential reduces the perturbation in the subspace of the center-of-mass wavenumber $k$ to:
\begin{equation}
\tilde{H}_k' = \sum_{s=-s_{\text{trun}}}^{s_{\text{trun}}} \left(V(s) + J_k \delta_{s,0}\right) \ket{k;s}\bra{k;s}
\end{equation}
Using this in Eq.~(\ref{eq:ScatteringMatrix}) yields for the $T$ matrix in the subspace $k$:
\begin{equation}
\tilde{T}_k(\omega) = \sum_{s,r=-s_{\text{trun}}}^{s_{\text{trun}}}\bra{k;s} \tilde{H}' \left(1-\tilde{G}_{0,k}(\omega) \tilde{H}' \right)^{-1} \ket{k;r}\ket{k;s}\bra{k;r},
\label{eq:scat_matrix}
\end{equation}
with 
\begin{equation}
\tilde{G}_{0,k}(\omega) = \sum_{s,r=-s_{\text{trun}}}^{s_{\text{trun}}}  \bra{k;s} G_0(\omega)\ket{k;r} \ket{k;s}\bra{k;r},
\label{eq:gok}
\end{equation}
i.e., the unperturbed Green function in the subspace $k$ ($G_{0,k}(\omega)$), projected onto the truncated $s$ basis. $G_{0,k}(\omega)$ can be obtained directly from Eqs.~(\ref{eq:Gzero_outside}) and (\ref{eq:Gzero}). Obviously, determining $\tilde{T}_k(\omega)$ involves manipulating matrices of dimension $2s_{\text{trun}} + 1$ only. Substituting $\tilde{T}_k(\omega)$ into Eq.~(\ref{eq:Greenfunction}) for the total Green function in the subspace $k$ yields
\begin{equation}
G_k(\omega) = G_{0,k}(\omega) + G_{0,k}(\omega) \tilde{T}_k(\omega) G_{0,k}(\omega),
\label{eq:gtotal}
\end{equation}
with $G_k(\omega) = \sum_{s,r} \bra{k;s} G(\omega)\ket{k;r} \ket{k;s}\bra{k;r}$ and $G_{0,k}(\omega) = \sum_{s,r} \bra{k;s} G_0(\omega)\ket{k;r} \ket{k;s}\bra{k;r}$. Thus the total Green function $G(\omega) = \sum_k G_k(\omega)$ may be obtained by manipulating matrices of dimension $2s_{\text{trun}} + 1$.

The total Green function can yield two different types of states, namely free and bound states. The free states correspond to the continuum part of the spectrum and are delocalized over all values of $s$. Bound states arise outside the continuum band at the poles of the scattering matrix. Bound states are localized around $s=0$ states and appear as sharp peaks in the density of states.\cite{economou_greens_1983}

It is well known from exciton theory that, for the situation where all transition dipole vectors are equal (as we consider), all oscillator strength resides in the $k=0$ band. This can be shown easily by considering the dipole operator in the $\ket{k;s}$ basis
\begin{equation}
\bra{g}\hat{\mu}\ket{k;s} = \mu \sum_n \bra{n;0} \sum_{m,s} e^{i\alpha_k s} e^{ik\left(m+s/2\right)} \ket{m;s} = \mu \delta_{s,0} \delta_{k,0}.
\end{equation}
Therefore, in the remainder of this paper, we focus solely on the $k=0$ band. The relevant parameters thus are $E_0$, $\epsilon_0 = \epsilon_{k=0}$, $\delta_0 = \delta_{k=0}$, $A$, and $\Gamma$. Henceforth, we set $E_0 = 0$ (without loss of generality) and express $\delta_{0}$, $A$, and $\Gamma$ in units of $\epsilon_{0} = |t_{\text{e}} + t_{\text{h}}|$, reducing the number of independent free parameters to three.

Alternatively to the Green function method presented above, one could directly diagonalize the Hamiltonian. This would require some form of basis truncation in order to keep it computationally feasible. The basis truncation will discretize the continuum part of the spectrum. Consequently, Hamiltonian diagonalization is suitable for obtaining the absorption spectrum only when the homogeneous linewidth is larger than the average energy separation between states inside the continuum part of the spectrum. In this large linewidth regime, the interpretation of the absorption spectrum becomes harder, since one can not properly distinguish bound and free states. The large homogeneous linewidth thus clouds the interpretation of the absorption spectrum. Instead, our Green function method is applicable for all values of the linewidth and therefore allows us to distinguish bound and free states and interpret the absorption spectrum in more detail.

%%%
%%% Results section
%%%
\section{\label{sec:results}Results}

\subsection{\label{sec:generalanalysis} General analysis of the absorption spectrum}
As we showed in Section~\ref{sec:GreenFunction}, calculating the Green function and, hence, the absorption spectrum using the scattering approach, involves the inversion of matrices of dimension $2s_{\text{trun}}+1$. This allows for an efficient numerical procedure to obtain these quantities upto values of $s_{\text{trun}}$ that are large enough to obtain spectra that are fully converged, i.e., that do not change anymore upon further increasing $s_{\text{trun}}$. In Appendix~\ref{sec:convergence}, we show that using $\Gamma = \epsilon_0/5$, the spectra are converged at $s_{\text{trun}} = 10$ already. Therefore, we use $s_{\text{trun}} = 15$ unless stated otherwise.

\begin{figure}[!htb]
	\includegraphics[width=8.5cm]{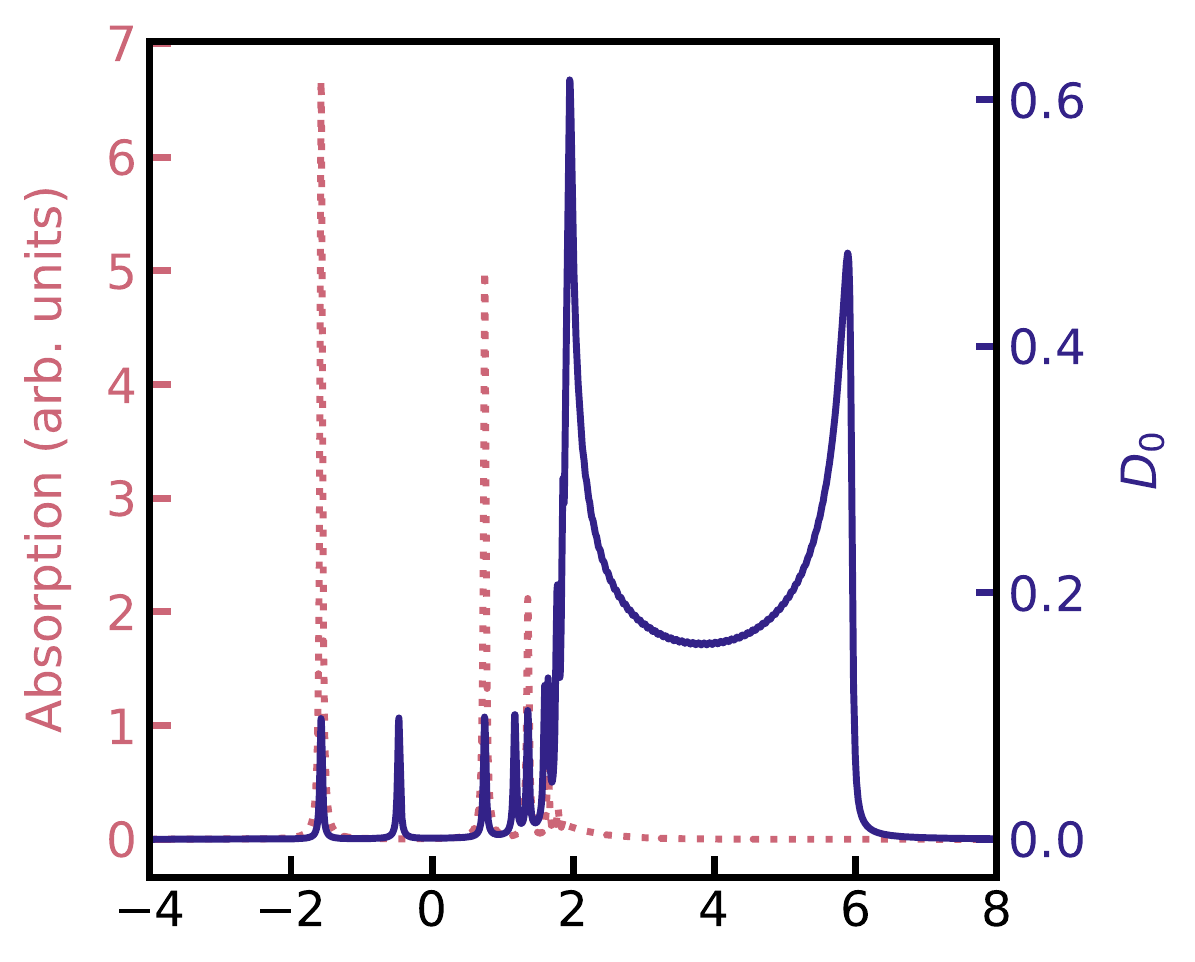}
	\caption{Density of states (solid blue) and absorption spectrum (dotted red) as calculated from Eqs.~(\ref{eq:DOS}) and (\ref{eq:absorption}) for the $k=0$ band, respectively. The parameters used are $\delta_0 = 0$, $A = 4\epsilon_0$, $\Gamma = 0.03\epsilon_0$, and $s_{\text{trun}} = 80$. The high $s_{\text{trun}}$ is needed to obtain a converged DOS. }
	\label{fig:DOS}
\end{figure}

In addition to the absorption spectrum, it is useful to consider the density of states (DOS) in the $k=0$ band, which is given by
\begin{equation}
 D_0(\omega) =  - \frac{1}{\pi} \text{Im}\left(\sum_s \bra{k=0;s}G(\omega)\ket{k=0;s}\right).
\label{eq:DOS}
\end{equation}   
A typical $D_0(\omega)$ is shown in Fig.~\ref{fig:DOS}, together with the corresponding absorption spectrum. The DOS shown is continuous in the range $2\epsilon_0 < \omega < 6\epsilon_0$, corresponding to the band of the unperturbed Green function $G_0$. This continuum part of the DOS, corresponding to free states, mostly follows the cosine dispersion of Eq.~(\ref{eq:oddenergies}) shifted over $J_{0}$. It shows van Hove singularities at both band edges \cite{economou_greens_1983, toyozawa_coexistence_1967} and is centered around $\delta_0 + A$.

To the left of the continuum band, sharp peaks with a FWHM of $2\Gamma $ appear, corresponding to bound states. The bound states have a finite charge separation due to the confining Coulomb well. The distribution of the bound states becomes more concentrated near the continuum band edge, as is typical for systems with a Coulomb potential, such as the hydrogen atom and GaAs. \cite{lee_wannier_1979}

Comparison of the DOS and absorption spectrum in Fig.~\ref{fig:DOS} shows that not all bound states have oscillator strength. This can be explained using the symmetry of the perturbation Hamiltonian $H'$. The eigenfunctions of $H'$ can be even or odd around the exciton position and only the even wave functions 
%{\color{red}can} 
have a non-zero exciton contribution, and hence, a nonzero transition dipole moment. The contribution of continuum states to the absorption spectrum is small, as there is relatively little mixing between exciton and CT states. At the end of this subsection we provide a more detailed analysis of the contribution of continuum states to the absorption spectrum.

\begin{figure*}[htb!]
	\includegraphics[width=17cm]{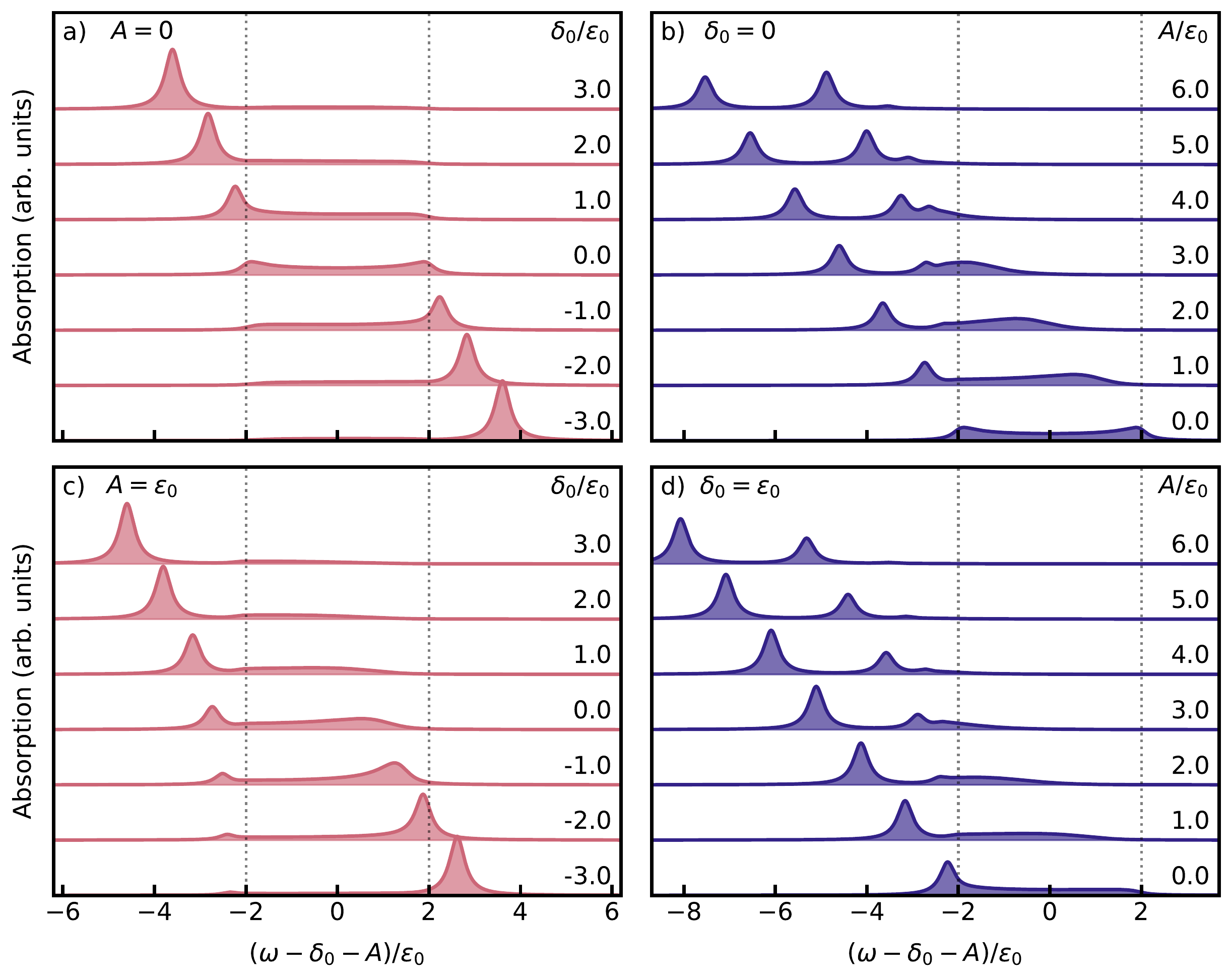}
	\caption{Absorption spectra calculated for varying $\delta_0$ (a, c) or $A$ (b, d) while keeping the other parameter fixed at the value indicated in the figure. The energy axis is chosen such that the continuum band is centered at zero. The dotted lines indicate the edges of the continuum band.}
	\label{fig:AbsRange}
\end{figure*}

The CT system parameterized by $\delta_0$ and $A$ can produce a wide range of different absorption spectra as shown in Fig.~\ref{fig:AbsRange}. First the effect of the detuning $\delta_0$ on the absorption spectrum is studied while neglecting the Coulomb well, i.e., using $A=0$ (Fig.~\ref{fig:AbsRange}a). This case reduces to a single-impurity problem because the potential $V(s) = \delta_{0}\delta_{s,0}$ then. This leads to an exact solution for the absorption spectrum, as described in {Appendix~\ref{sec:SingleImpurity}}. A single impurity induces exactly one bound state, for which the energy can be expressed analytically as (see {\mbox{Appendix}~\ref{sec:SingleImpurity}}): 
\begin{equation}
E_{\text{bound}} = \delta_0 + A - \text{sgn}(\delta_0+A) \sqrt{(\delta_0 + A)^2 + 4\epsilon_0^2},
\label{eq:boundEnergy}
\end{equation}
where $\text{sgn}(\delta_0+A)$ is the sign ($+$ or $-$) of $\delta_0 + A$. In the limit of $\epsilon_0 \ll \left|\delta_0 + A\right|$ we find $E_{\text{bound}} \approx 0$ and there is no contribution of continuum states to the spectrum. In the opposite limit $\epsilon_0 \gg \left|\delta_0+A\right|$ we have $E_{\text{bound}} \approx \pm 2\epsilon_0$ and the notion of bound states disappears while continuum states possess most oscillator strength. These limits lead to the definition of the coupling strength between the excitons and the CT states as the ratio
\begin{equation}
M = \frac{2\epsilon_0}{\left|\delta_0 + A\right|}.
\label{eq:CouplingStrength}
\end{equation}

The further the bound state moves away from the continuum, the more oscillator strength it accumulates at the expense of the continuum part of the spectrum, as is clear from Fig.~\ref{fig:AbsRange}. The mixing of exciton and CT states is highest when $\delta_0 = 0$, which is the case when the resonant excitation transfer coupling $J_0$ equals the exciton-CT energy gap $\Delta$, i.e., the exciton state becomes resonant with the nearest-neighbor CT state. Since $\Delta$ is generally positive, the former condition is mostly met when $J_0 > 0$, i.e., for H-aggregates.

In Fig.~\ref{fig:AbsRange}b the Coulomb well depth $A$ is varied while the detuning $\delta_0$ is set to zero. Since the Coulomb potential will always have a positive contribution to the energy, we only consider positive values of $A$. As expected, increasing $A$ leads to the formation of additional bound states outside the band edge. When $A$ becomes much larger than $\epsilon_0$, the rate at which bound states arise increases. However, the newly formed bound states have little oscillator strength and thus will hardly be visible in the absorption spectrum. It depends on the sign of $\delta_0$ at which band edge the bound states form, as can be seen in Fig.~\ref{fig:AbsRange}c. For large Coulomb well depths, e.g., $A > 4.0 \epsilon_0$, we approach the limit where the absorption spectrum is dominated by the Frenkel exciton state and the $|s|=1$ CT states, leading to two dominant absorption peaks (Figs.~\ref{fig:AbsRange}b and \ref{fig:AbsRange}d). 
%{\color{gray} [REWRITE, this section needs a better mathematical definition of resonance states]
%In principle there are $2s_{\text{trun}}+1$ bound states allowed by our method due to the size of $H'$, however, the actual number of bound states discernible in the DOS may be much lower. The formation of bound states is governed by the real part of the unperturbed Green's function. This can be seen from Eq.~(\ref{eq:Tgeometric}) where the denominator can only produce singularities when $G_0$ is purely real. Inside the band, $G_0$ will always maintain an imaginary part. Bound states can therefore only form outside the band when the Green's function equals the inverse energy of a perturbed state. 
%
%Resonance eigenstates occur when the denominator of the $T$ matrix becomes very small. Through the addition of bound states the imaginary part of the Green's function within the continuum band will diminish and thus allowing the formation of resonance eigenstates. The resonance eigenstates are partially confined to the exciton and therefore possess more oscillator strength than free states. The corresponding eigenenergy of the resonance eigenstates determines the shape of the absorption spectrum within the continuum band. As the resonance energies tend towards the bound states, e.g. left band edge, the continuum part of the spectrum is seen to accumulate a positive skew.
%
%[END REWRITE]}

\begin{figure}
	\includegraphics[width=8.5cm]{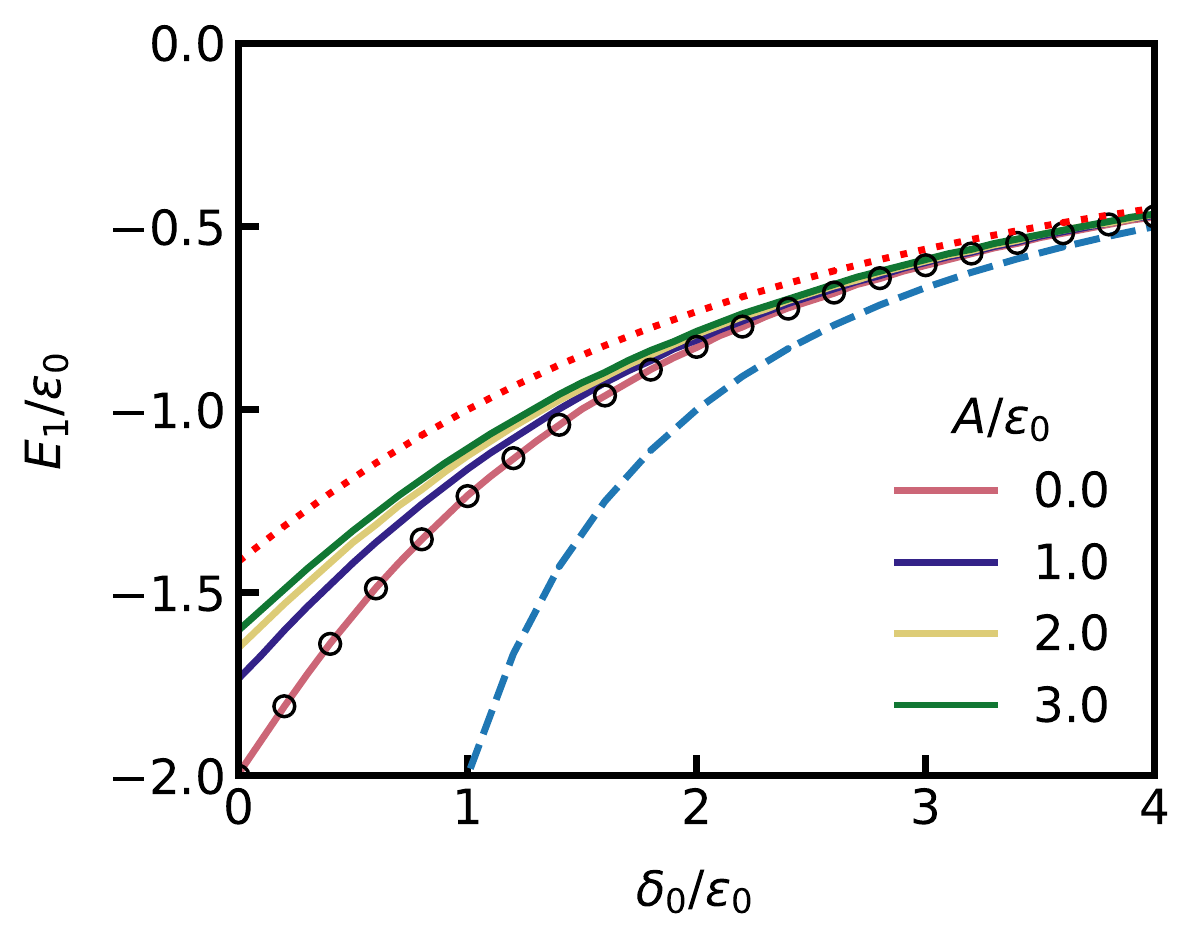}
	\caption{Numerically obtained position $E_1$ of the lowest-energy peak in the absorption spectrum as function of $\delta_0$ for different values of $A$ (solid curves). The open black circles indicate the analytic result of Eq.~(\ref{eq:boundEnergy}) for $A=0$. The dashed blue line indicates the perturbative solution of Eq.~(\ref{eq:perturbative}), while the red dotted line corresponds to the nearest-neighbor solution of Eq.~(\ref{eq:nearestNeighbor}).}
	\label{fig:lowestpeak}
\end{figure}

Both $\delta_0$ and $A$ have an effect on the energy $E_1$ of the first bound state as is shown in Fig.~\ref{fig:lowestpeak}. When both $\delta_0$ and $A$ are zero, $E_1$ is seen to lie at the left band edge. The value plotted is shifted slightly due to the finite linewidth, which causes the peak maximum to lie just next to the band edge. Upon increasing $\delta_0$, $E_1$ tends towards the energy of the $k=0$ Frenkel exciton. Setting $A=0$ is effectively equal to truncating the Coulomb well beyond nearest neighbors ($s_{\text{trun}}=0$) and therefore, in Fig.~\ref{fig:AbsRange}a and Fig.~\ref{fig:lowestpeak}, we see perfect agreement between the $A=0$ curve and the result of Eq.~(\ref{eq:boundEnergy}).

As observed in Fig.~\ref{fig:lowestpeak}, using a finite Coulomb well depth $A$, causes $E_1$ to shift upward. The reason is that upon increasing $A$, the continuum band (centered at $\omega = \delta_0 + A$) shifts to higher energies, pulling the bound states with it. For $A >3\epsilon_0$ the energy separation between the bound states becomes much larger than the CT coupling $\epsilon_0$ and the peak position converges. At this point one approaches the regime where only the nearest-neighbor CT state affects the exciton. One can show that in this limit the energy of the lowest bound state is given by \cite{hestand_polarized_2015}
\begin{equation}
E_1 = \frac{\delta_0}{2} - \sqrt{\left(\frac{\delta_0}{2}\right)^2 + \epsilon_0^2},
\label{eq:nearestNeighbor}
\end{equation}
a result that is shown by the dotted red line in Fig.~\ref{fig:lowestpeak}. The oscillator strength ratio between the first and second peak is then $\left(\delta_0^2 + 4\epsilon_0^2 + 2\delta_0 \sqrt{\delta_0^2 + 4\epsilon_0^2}\right)/\left(\delta_0^2 + 4\epsilon_0^2 - 2\delta_0 \sqrt{\delta_0^2 + 4\epsilon_0^2}\right)$. Thus, an increase of the detuning shifts oscillator strength towards the first peak as can be seen when comparing Figs.~\ref{fig:AbsRange}b and \ref{fig:AbsRange}d for the values of $A>5\epsilon_0$.

In the limit of large detuning the energy of the bound state can be described using first-order perturbation theory in the coupling $\epsilon_0$. By writing an effective Hamiltonian one can incorporate the effect of CT states as an effective resonance coupling coupling $J_{\text{CT}}$ between neighboring sites and an energy correction term $\Delta E_{\text{CT}}$, given by Ref.~\onlinecite{yamagata_designing_2012}:
\begin{equation}
J_{\text{CT}} = -2\frac{t_{\text{e}} t_{\text{h}}}{\Delta},
\end{equation}
and
\begin{equation}
\Delta E_{\text{CT}} = -2\frac{t_{\text{e}}^2 + t_{\text{h}}^2}{\Delta},
\end{equation}
respectively. The energy of the first peak is then given by
\begin{equation}
E_1 = E_0 + \Delta E_{\text{CT}} + 2(J_{\text{CT}}+J_0).
\label{eq:perturbative}
\end{equation}
Equation~(\ref{eq:perturbative}) is plotted as blue dashed line in Fig.~\ref{fig:lowestpeak}, where it becomes clear that this result only holds when the detuning $\delta_0$ is much larger than the bandwidth of the unperturbed Green function $2\epsilon_0$.

Figure~\ref{fig:lowestpeak} shows large differences between the various analytical approximations. The single-impurity result of Eq.~(\ref{eq:boundEnergy}) bridges the gap between the previously found nearest-neighbor and perturbative results of Eqs.~(\ref{eq:nearestNeighbor}) and (\ref{eq:perturbative}) respectively. Note that including a non-zero Coulomb well depth to Eq.~(\ref{eq:boundEnergy}) simply shifts the curve to the left, as  $\delta_0+A$ is the important quantity. Figure~\ref{fig:lowestpeak} also shows that the effect of the Coulomb well depth $A$ on the position of the lowest-energy peak can be quite significant.

From Fig.~\ref{fig:lowestpeak} one can infer the effect on the lowest-energy peak upon truncating the maximum charge separation in the basis set too much. Truncating the maximum charge separation between the electron and hole effectively creates an infinite potential at the truncation distance, thus increasing the effective Coulomb well depth. This results in a blueshift of the lowest-energy peak $E_1$. If one is not aware of this effect, one could wrongly include the truncation induced blueshift in the gas-to-crystal shift when interpreting experimental spectra. 
%{ \color{gray} [COMMENT] To make this point more clear we could create a figure like Fig 4 but as function of $s_{\text{trun}}$ with $A\sim 5\epsilon_0$. We could also just leave the above paragraph out. [END COMMENT]}

While for non-zero values of $\delta_0$ and $A$ most oscillator strength resides in the bound states, some of the continuum states may still contribute significantly to the absorption spectrum. For the single-impurity case ($A=0$), the relative continuum contribution to the spectrum is found in Appendix~\ref{sec:SingleImpurity} to be
\begin{equation}
Y_{\text{cont}} = \frac{\int d\omega I_{\text{cont}}(\omega)}{\int d\omega I(\omega)} = 1- \frac{\delta_0 }{\sqrt{\delta_0^2 + 4\epsilon_0^2}} \quad \left[A=0\right].
\label{eq:ContinuumAbs}
\end{equation}
When both the detuning $\delta_0$ and Coulomb well depth $A$ are zero, there are no bound states and $Y_{\text{cont}}=1$. Upon increasing the detuning, $Y_{\text{cont}}$ decreases and eventually approaches zero when $\delta_0 \gg \epsilon_0$ (and thus $M\ll\frac{1}{2}$), which means the absorption spectrum is dominated by bound states.

\begin{figure}
	\centering   
	\includegraphics[width=8.5cm]{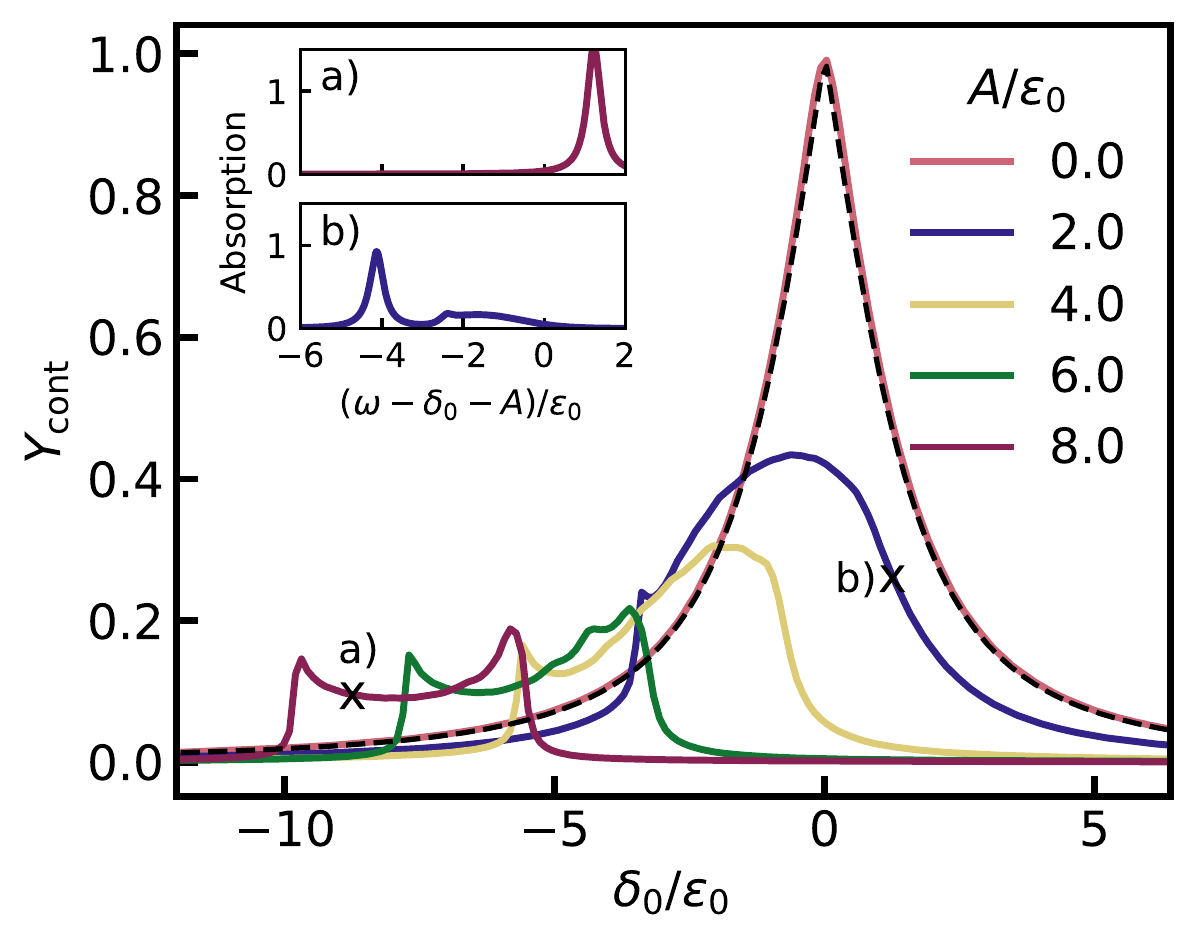}
	\caption{Relative contribution of continuum states to the absorption spectrum, $Y_{\text{cont}}$, as function of $\delta_0$ for various Coulomb well depths $A$. For small $A$ the continuum contribution is well described by the single-impurity results of Eq.~(\ref{eq:boundEnergy}) (dashed line). Upon increasing $A$, the contribution of the continuum states decreases and its dependence on $\delta_0$ reflects the DOS in the $k=0$ band at large charge separation. The inset shows the corresponding absorption spectrum at the points indicated. For the main plot we used $\Gamma = \epsilon_0/100$, and for the insets $\Gamma = \epsilon_0/5$.}
	\label{fig:Continuum}
\end{figure}

More generally, for a non-zero Coulomb well depth, the relative contribution of the bound states, $Y_{\text{bound}} = 1- Y_{\text{cont}}$, can be estimated numerically by integrating the positive part of the scattering term $G_0 T G_0$. As bound states are newly formed scattering states, they manifest themselves as positive contributions in $G_0TG_0$, while the diminished contribution of the continuum part of the spectrum is negative in sign. The relative weight of the continuum part of the spectrum is then obtained through $Y_{\text{cont}} = 1 - Y_{\text{bound}}$. Plotting the thus obtained $Y_{\text{cont}}$ as function of detuning and Coulomb well depth (see Fig.~\ref{fig:Continuum}) reveals a resonance condition of the bound states with the continuum. For a shallow Coulomb well we recover the result of Eq.~(\ref{eq:ContinuumAbs}). For large Coulomb well depths ($A > 4 \epsilon_0$) the dependence of $Y_{\text{cont}}$ on $\delta_0$ converges and probes the density of states of the $k=0$ band at high charge separation. As the states at high charge separation do not feel the Coulomb attraction, the dispersion of these states remains a cosine centered at $E_{CT}$ and with a bandwidth of $4\epsilon_0$. 
%At high charge separation, $s \rightarrow \infty$, the energy of states in the diabatic basis (Eq.~\ref{eq:comtrans}) approaches $E_{CT}$ while the coupling remains constant at $\epsilon_0$ and, hence, the DOS follows a cosine dispersion centered at $E_{CT}$ and a bandwidth of $4\epsilon_0$. 
In the limit of a large Coulomb well $A$ and a large negative detuning $\delta_0$ the Frenkel exciton becomes resonant with CT states with a large charge separation. As the density of states is very high at large charge separation, this resonance causes a Fano effect,\cite{fano_effects_1961, kolata_molecular_2014, monahan_charge_2015} which is manifested by a sharp peak in the absorption spectrum. As seen in the inset of Fig.~\ref{fig:Continuum}, such a Fano resonance will only occur when $J_0 \sim \Delta + A$.

The above observations highly depend on the relative size of the CT-coupling, $\epsilon_0 = t_{\text{e}} + t_{\text{h}}$, with respect to the detuning $\delta_0$ and the Coulomb well depth $A$. Previous research \cite{hestand_interference_2015} has shown that the sign of $t_{\text{e}}$ and $t_{\text{h}}$ is very sensitive to the molecular packing and, thus, $t_{\text{e}}$ and $t_{\text{h}}$ may interfere destructively (out of phase) or constructively (in phase). As a result the ratios $\delta_0 / \epsilon_0$ and $A/\epsilon_0$ are small/large when the CT integrals are in/out of phase.

\subsection{Two-peak nature of the absorption spectrum}

\begin{figure}
	\centering
	\includegraphics[width=8.5cm]{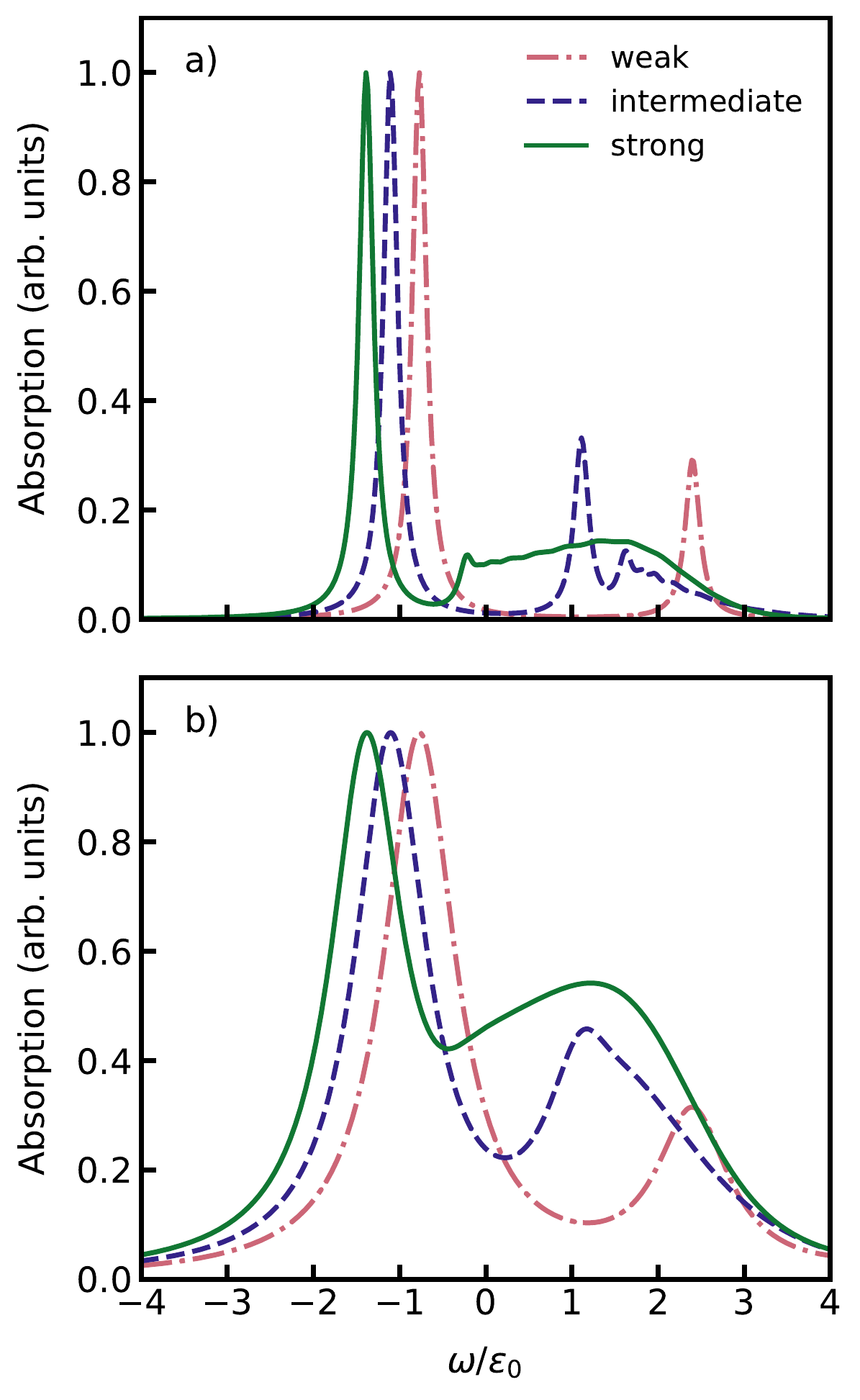}
	\caption{Absorption spectrum for weak ($M=1/4$), intermediate ($M=1/2$), and strong ($M=1$) coupling where $M$ is defined in Eq.~(\ref{eq:CouplingStrength}). In (a) the homogeneous linewidth is small ($\Gamma = \epsilon_0/10$) leading to many different peaks in each coupling regime. In (b) a bigger linewidth is used ($\Gamma = \epsilon_0/2$) and the details which were first visible in (a) have now blended together into just two peaks. In both figures $A = 3 \delta_0$. For easy comparison, all spectra are normalized based on the maximum peak height.}
	\label{fig:Gamma}
\end{figure}

Earlier experimental and theoretical works on this topic have noted that CT couplings often lead to spectra with only two peaks.\cite{oleson_perylene_2019, hestand_interference_2015, lalov_vibronic_2007, petelenz_charge-transfer_2015} Using the Green function method described above, the nature of these two peaks can now be investigated in more detail. In Fig.~\ref{fig:Gamma} absorption spectra for different coupling regimes are compared, using either a small or a large homogeneous linewidth. For a small linewidth a collection of separate peaks is observed for all coupling strengths. Upon increasing the homogeneous linewidth, the peaks will merge and the aforementioned two-peak spectrum is obtained. The lowest-energy peak is always due to a single bound state and hence has a simple Lorentzian lineshape. The second peak, however, is very different in nature for weak, intermediate, and strong coupling, respectively, as will be analyzed below.

In the weak-coupling limit ($M\ll1/2$) two dominant bound states are visible in the absorption spectrum, which result from mixing between the Frenkel exciton and nearest-neighbor CT states. Upon an increase of the homogeneous linewidth $\Gamma$ each peak still corresponds to a bound state of the system and only the FWHM of the two peaks increases. For intermediate coupling ($M \approx 1/2$) and large $\Gamma$ the higher-energy peak is formed by multiple bound states together with a small contribution of continuum states. These peaks are close in energy and the second peak no longer has a Lorentzian lineshape but is now skewed towards higher energies. Finally, for strong coupling ($M \gg 1/2$) the second peak is dominated by free states causing its position to be close to the high-energy edge of the continuum band. The continuum states fill the region between the bound state and the high-energy continuum band edge. The second peak now skews towards lower energies.

\begin{figure}
	\includegraphics[width=8.5cm]{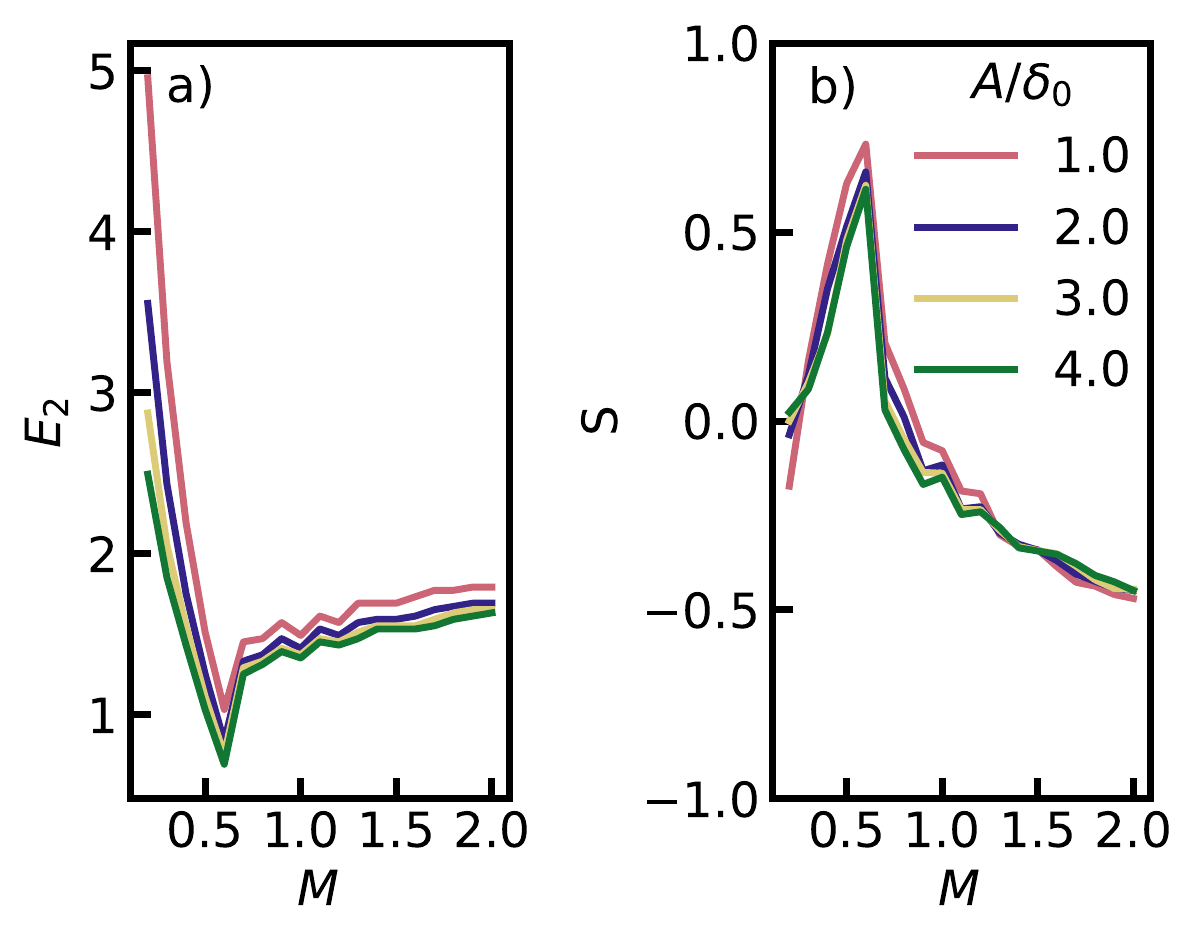}
	\caption{Position $E_2$ (a) and skewness $S$ (b) of the higher-energy peak in the absorption spectrum as function of the coupling strength $M$ for several values of $A/\delta_0$.}
	\label{fig:HWHM}
\end{figure}

To provide a more quantitative picture we analyze the second peak separately. The first peak is excluded from the spectrum by fitting it with a Lorentzian lineshape and subtracting this peak. The residual spectrum is then analyzed in terms of the energy $E_2$ at which it reaches its maximum and a measure of the peak's skewness defined as
\begin{equation}
S = \frac{\text{RWHM}-\text{LWHM}}{\text{RWHM}+\text{LWHM}},
\end{equation}
where $\text{RWHM}$ and $\text{LWHM}$ refer to the right and left width at half maximum respectively. The results for $E_2$ and $S$ are shown in Fig.~\ref{fig:HWHM} as a function of $M$ for several values of $A/\delta_0$. Most notably we see a sharp transition in the peak position of the second peak when $M=1/2$. For small $M$ the energy of the second peak $E_2$ lies at the high-energy side of the continuum band of $G_0$. Upon increasing the detuning and Coulomb well depth, $E_2$ moves towards the lower-energy edge of the continuum band of $G_0$. Then a sharp transition is visible after which $E_2$ suddenly increases again. This point determines the transition from the strong- to the weak-coupling limit. Around this point a second bound state will appear in the spectrum which quickly dominates the residue spectrum. 

Looking at the LWHM and RWHM underlines this interpretation. For a high $M$ the residue spectrum is mostly skewed towards the left as indicated by the negative skewness. After the transition point a second bound state forms and both the LWHM and RWHM quickly converge to $\Gamma$, the half-width at half maximum of the Lorentzian peak associated with this bound state. As a result the skewness $S$ becomes zero. 

The energy and skewness of the second peak may serve as practical measure to estimate the CT coupling strength $M$. Future studies will have to demonstrate how vibronic coupling, which adds high-energy vibronic absorption bands, affects this measure.

\begin{figure}
	\includegraphics[width = 8.5cm]{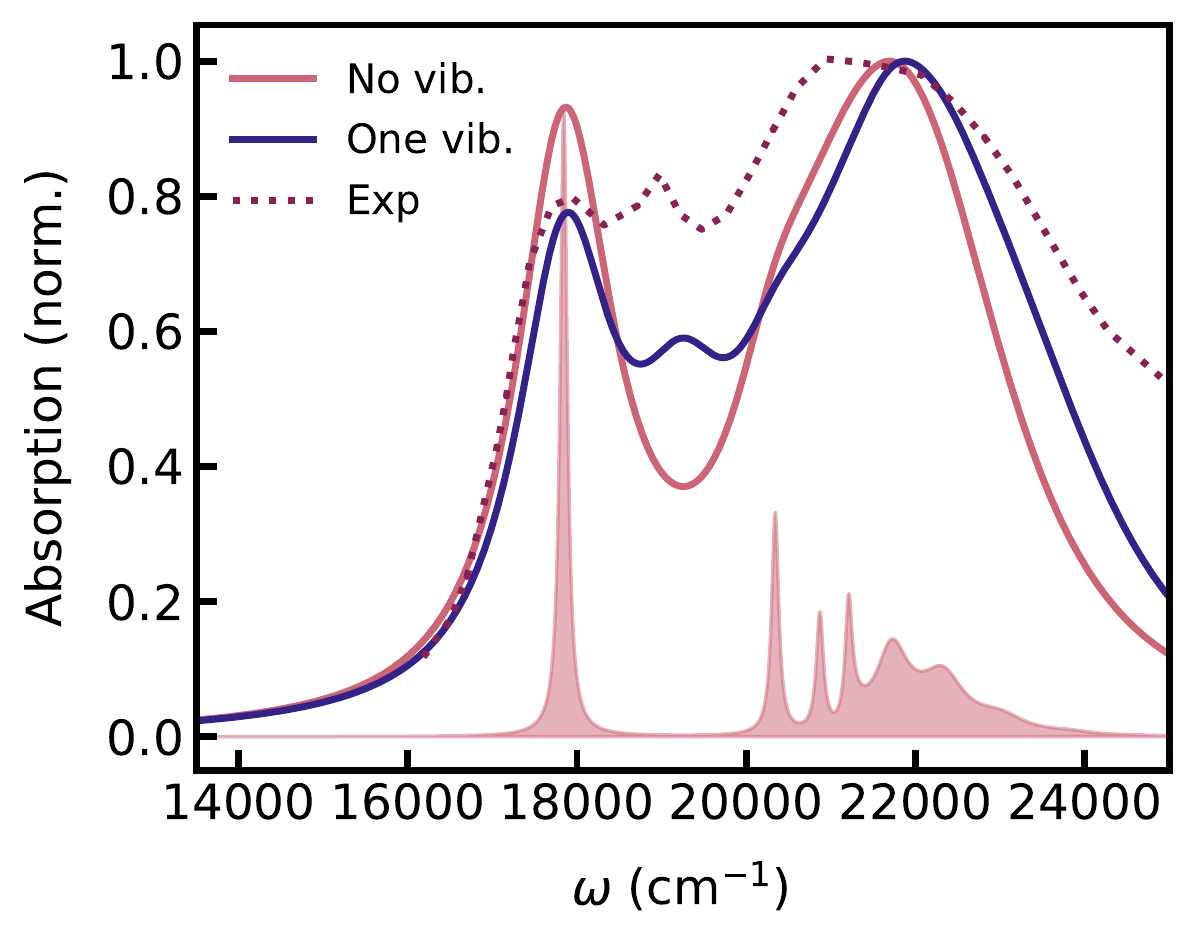}
	\caption{Experimental absorption spectrum of PDI (red dots) digitally extracted from Ref.~\onlinecite{oleson_perylene_2019} and the fit to this spectrum using the model and Green function method described in Section~\ref{sec:model} (solid red curve). The light red curve shows the same spectrum as the red solid curve but with a small linewidth which reveals multiple bound states and a skewed continuum contribution. The blue solid line gives the fit with a vibration added hand-wavingly by shifting a copy of the origional spectrum by a vibrational quantum and weighting it by a Franck-Condon factor. The parameters used are $E_0 = 18260 \textrm{ cm}^{-1}$, $\epsilon_0 = t_\text{e} + t_\text{h} = 1386\textrm{ cm}^{-1}$, $J_0 = \sum_n J_{0,n}/\epsilon_r = 547.27 \textrm{ cm}^{-1}$, $\Delta = 950 \textrm{ cm}^{-1}$, $A = 4000 \textrm{ cm}^{-1}$, and $\Gamma = 700 \textrm{ cm}^{-1}$. 
		%The resonant excitation transfer coupling $J_0$ is obtained by summing the values in Table 1 of Ref.~\onlinecite{oleson_perylene_2019} divided by the dielectric constant given in Table S1. 
	The entire spectrum is shifted by $\Delta E = 1000 \textrm{ cm}^{-1}$ to align the lowest-energy peak. For the vibronic replica a Franck Condon factor of $\text{FC} = \lambda e^{-\lambda^2/2}$ is used with a Huang-Rhys factor $\lambda^2=0.6$ and a vibrational frequency of $\omega_{\text{vib}} = 1400 \textrm{ cm}^{-1}$. 
	%{\color{red} The monomer energy $E_0$, CT couplings $t_\text{e}$ and $t_\text{h}$ (without scaling), resonant-exciton-transfer couplings $J_{0,n}$, dielectric constant $\epsilon_r$, and Huang-Rhys factor $\text{HR}$ are taken from Ref.~\onlinecite{oleson_perylene_2019}. The other parameters are taken as free parameters such that the best fit is obtained.  }
	All parameters where taken from Ref.~\onlinecite{oleson_perylene_2019}, except for the energy shift $\Delta E$, the detuning $\Delta$, and the Coulomb well depth $A$, which have been taken as free fitting parameters. By contrast to Ref.~\onlinecite{oleson_perylene_2019}, the CT couplings $t_{\text{e}}$ and $t_{\text{h}}$ were not scaled relative to their ab initio values.
	}
	\label{fig:OlesonFit}
\end{figure}

To illustrate the potential of the presented Green's function method we fit the spectrum measured for perylene-diimide aggregates, \cite{oleson_perylene_2019} shown in Fig.~\ref{fig:OlesonFit}. The resonant exciton transfer interactions $J$ and the CT couplings have been taken equal to the ab-initio values. \cite{oleson_perylene_2019} In order to reach a good fit in Ref.~\onlinecite{oleson_perylene_2019}, the ab-initio values for $t_{\text{e}} = 994 \textrm{ cm}^{-1}$ and $t_{\text{h}} = 392 \textrm{ cm}^{-1}$ had to be scaled with an additional fitting parameter. In our case we do not need this additional scaling. Just as in Ref.~\onlinecite{oleson_perylene_2019} the detuning and the Coulomb well depth are taken as free fitting parameters. Since our model does not allow for vibronic coupling as of yet, we neglect those for now. Instead, we add a vibration in a hand-waving fashion by creating a copy of the absorption spectrum weighted by the relevant Franck-Condon factor and shifted by a vibrational quantum, both taken from Ref.~\onlinecite{oleson_perylene_2019}. The resulting spectrum with one vibrational mode that for varying $\delta_0$ and $A$ best resembles the experimental one, is shown as blue solid curve in Fig.~\ref{fig:OlesonFit} (the red solid curve is the spectrum calculated without a vibrational replica). 

The resulting fits reproduce a (mainly) two-peak spectrum with the correct energy splitting and peak height ratio between the two main peaks. The fit with one vibration gives an impression of how the fit can be improved when including vibronic coupling. The proper addition of vibronic coupling will create a larger tail on the right side of the high-energy peak and increase the oscillator strength of the bound states around $21000 \text{ cm}^{-1}$, as described by Ref.~\onlinecite{hestand_extended-charge-transfer_2016}. The fits shown in Fig.~\ref{fig:OlesonFit} are obtained using an intermediate coupling strength $M=0.63$, which indicates that the second peak is formed by a mixture of bound and continuum states. 
 
We perform the same analysis as done for Fig.~\ref{fig:HWHM} directly for the experimental spectrum (red dotted line in Fig.~\ref{fig:OlesonFit}) and find the skewness of the second peak to be $S=0.2$. Comparing this value with Fig.~\ref{fig:HWHM} suggests an intermediate coupling strength $M$, which agrees with our fit. We conclude that an intermediate coupling strength $M$ may be applicable to PDI aggregates and that the skewness of the second peak can be a good indicator for the exciton-CT coupling strength $M$. 

Our fitting procedure using the Green function technique does not provide a qualitatively better fit than the one obtained in Ref.~\onlinecite{oleson_perylene_2019}. It does, however, result in different values for the Coulomb well depth $A$ and detuning $\Delta$, leading to an intermediate coupling regime. This means one has to be careful when truncating the basis to nearest-neighbor CT states as done in Ref.~\onlinecite{oleson_perylene_2019}, since it limits the parameter space and could underestimate the contribution of several bound states and continuum states. As mentioned already, further studies will have to show how proper inclusion of vibronic effects influence these findings.

%%%
%%% Conclusions section
%%%
\section{Conclusions}
\label{sec:Conclusion}
In this paper we give insight into the exciton and CT states that constitute the absorption spectrum of organic molecular aggregates. The empirical observation of a two peak absorption spectrum is confirmed in the limit where the linewidth $\Gamma$ is larger than the average energy spacing between states. We have found that the two-peak spectrum occurs regardless of the coupling strength between exciton and CT states, but is rather a feature of the large homogeneous linewidth. In particular, the second peak often results from a collection of bound and free states (see Fig.~\ref{fig:Gamma}), which makes it hard to interpret experimental fits. 

Both the detuning $\delta_0$ and Coulomb well depth $A$ play an important role in the position of the first peak (see Fig.~\ref{fig:lowestpeak}). By truncating the range of the Coulomb well, a new analytical expression is found for the position of the lowest-energy peak, which bridges the gap between previous analytical results (see Eq.~(\ref{eq:boundEnergy})). 
%Compared to methods that truncate the charge separation, the effect of the Coulomb well on the lowest-energy peak may be falsely attributed to a gas-to-crystal shift.

Using the position and lineshape of the higher-energy peak, we can estimate the coupling strength between the exciton and CT states (see Fig.~\ref{fig:HWHM}), which is determined by the ratio of CT couplings with  the Coulomb well energy, $M = \frac{2\epsilon_0}{\left|\delta_0 +A\right|}$. This provides a new way to obtain model parameters from the experimental absorption spectrum alone.

Previous numerical methods are prevalent mostly in the weak-coupling regime. Hence, the method presented here provides a new approach to understand experimental spectra in the intermediate- and strong-coupling limit, while avoiding costly numerical calculations. While this research bridges the gap between earlier methods, there are limitations which remain to be solved. Examples are the lack of vibronic coupling and the use of a relatively simple system-bath coupling. As indicated by Lalov et. al., \cite{lalov_vibronic_2007} it is possible to include vibronic coupling in a Green's function formalism, which provides a perspective for future improvements. 

%Another limitation of the current theory is the inclusion of non-Markovian bath interactions. The complex self energy $i\Gamma$ performs reasonably well for exciton systems where dynamic disorder is uncorrelated on each site. However, for CT states the dynamic energy fluctuations may not be independent for each state, making the linewidth energy dependent.

It is possible to extend the model to 2D and 3D systems and still obtain analytical expressions for $G_0(\omega)$. Extending the number of spatial dimensions also alters the effect of the scattering matrix $T$. As is well known from scattering theory, \cite{shen_quantum_1997, economou_greens_1983} a perturbation will always lead to at least one bound state in 1D and 2D systems but not necessarily in 3D systems.

%%%
%%% Acknowledgements
%%%
\begin{acknowledgments}
The authors are grateful for the discussions with dr. T. Kunsel.
\end{acknowledgments}

%%%
%%% Data availability
%%%
\section*{Data Availability Statement}
The data that support the findings of this study are openly available in ImpurityGreenFunction at https://doi.org/10.6084/m9.figshare.19448252.v1, reference number 19448252.

%%%
%%% appendices
%%% 
\appendix

%%%
%%% Convergence
%%%
\section{Convergence when increasing $s_{\text{trun}}$}
\label{sec:convergence}
\begin{figure}[htb!]
	\centering
	\includegraphics[width=8.5cm]{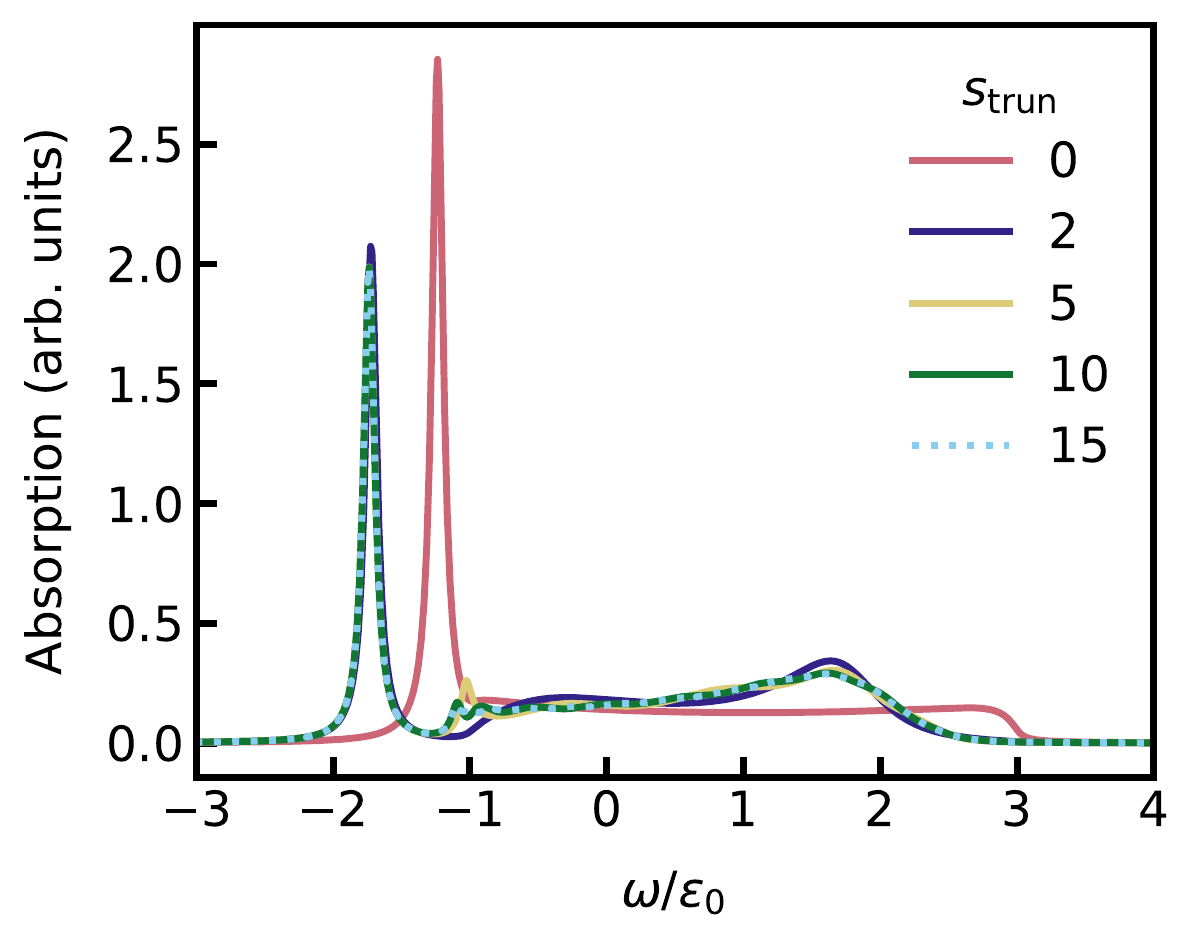}
	\caption{Absorption spectrum for different values of $s_{\text{trun}}$. The spectrum shows convergence around $s_{\text{trun}}=10$. Parameters used are $\delta_0 = 0$, $A = \epsilon_0$, and $\Gamma = 0.05\epsilon_0$.}
	\label{fig:Convergence}
\end{figure}

The Green function method used in this paper circumvents truncation of the basis by truncating the Coulomb potential instead. In Fig.~\ref{fig:Convergence} the effect of this truncation on the absorption spectrum is shown for a shallow Coulomb well ($A=\epsilon_0$). When $s_{\text{trun}}=0$, the Coulomb potential is effectively neglected and only an energetic detuning between the exciton state and the CT states is included. This results in an absorption spectrum where both the bound state and continuum contributions are significant. When including the effect of the Coulomb well, by increasing $s_{\text{trun}}$, the dominant bound state moves away from the continuum and new bound states form around the lower-energy band edge. The Green function accumulates a real part within the continuum part, causing the overall shape of the absorption spectrum to change. The spectra at $s_{\text{trun}} = 10$ and $s_{\text{trun}} = 15$ barely show any differences, hence we use $s_{\text{trun}} = 15$ as we found all shown absorption spectra to be converged at this value.

%%%
%%% Analytical bound state results
%%%
\section{Absorption spectrum for $s_{\text{trun}} = 0$}
\label{sec:SingleImpurity}

When the Coulomb well range is truncated using $s_{\text{trun}} = 0$, the Green function can be found analytically. In this particular case all site-basis states have an energy of $E_{\text{CT}}$, while the $k=0$ exciton state has an energy of $E_{\text{CT}} - A - \delta_0$. In this case all scattering happens at the exciton state ($s=0$), making the system equivalent to a single-impurity system. For a single impurity the bound state energy in the $k=0$ band is simply computed by determining the poles of the scattering matrix $T$ determined by
\begin{equation}
1-\left(\delta_0 + A\right) G_0(E_{\text{bound}}) = 0.
\end{equation}
This equation is readily solved to yield
\begin{equation}
E_{\text{bound}} = E_{\text{CT}} - \text{sgn}(\delta_0 + A) \sqrt{(\delta_0 + A)^2 + 4\epsilon_0^2} 
\label{eq:bound_energy}
\end{equation}
where $\text{sgn}$ is the sign ($+$ or $-$) of $\delta_0 + A$.
By evaluating the residue of $G$ at $E_{\text{bound}}$ we obtain for the normalized wave function $\ket{b_0}$ of the bound state $\braket{k=0;s|b_0} = \beta^{|s|}\alpha$, with
\begin{equation}
\beta =\frac{-\delta_0 - A - \text{sgn} (\delta_0 + A) \sqrt{(\delta_0 +A)^2 + 4\epsilon_0^2}}{2\epsilon_0},
\label{eq:beta}
\end{equation}
and
\begin{equation}
\alpha= \left(\frac{\delta_0 + A}{\sqrt{(\delta_0 + A)^2+4\epsilon_0^2}}\right)^{1/2}.
\label{eq:alpha}
\end{equation}
The absorption spectrum of the bound state is given by a Lorentzian lineshape centered around $E_{\text{bound}}$ and height $\alpha^2$,
\begin{equation}
I_{\text{bound}}(\omega) = \frac{1}{\pi}\frac{\alpha^2 \Gamma}{(\omega-E_{\text{bound}})^2+\Gamma^2}.
\label{eq:Ibound}
\end{equation}
From Eqs.~(\ref{eq:beta}) and (\ref{eq:alpha}), the average charge separation in the bound state is found to be
\begin{equation}
\braket{s}_{\text{bound}} = \frac{2\beta^2}{1+\beta^4}.
\end{equation}

In the case of a single impurity, the continuum part of the absorption spectrum can also be found by inserting Eqs.~(\ref{eq:scat_matrix})-(\ref{eq:gtotal}) into Eq.~(\ref{eq:absorption}) and evaluating the imaginary part within the continuum band of $G_0$. This leads to
\begin{equation}
	I_{\text{cont}}(\omega) = \frac{1}{\pi}\frac{\sqrt{4\epsilon_0^2-\left(\omega-\delta_0 -A\right)^2}}{4\epsilon_o^2-\omega^2+2\omega(\delta_0+A)} \text{ for } \left|\frac{\omega-\delta_0-A}{2\epsilon_0}\right|<1.
	\label{eq:cont_osc}
\end{equation}
Equations~(\ref{eq:bound_energy})-(\ref{eq:alpha}) and (\ref{eq:cont_osc}) reproduce the the absorption spectra in Fig.~\ref{fig:AbsRange}a. We find the total contribution of the bound state and continuum states to the absorption spectrum by integrating Eq.~(\ref{eq:Ibound}) and (\ref{eq:cont_osc}) respectively. 
The contribution of the bound state is
\begin{equation}
Y_{\text{bound}} = \frac{\int d\omega I_{\text{bound}}(\omega)}{\int d\omega I(\omega)} = \left|\braket{k=0;s=0|b_0}\right|^2 =  \alpha^2.
\end{equation}
By integrating Eq.~(\ref{eq:cont_osc}) or using the fact that $Y_{\text{bound}} + Y_{\text{cont}} = 1$ we find the total contribution of continuum states to the absorption spectrum
\begin{equation}
Y_{\text{cont}} = \frac{\int d\omega I_{\text{cont}}(\omega)}{\int d\omega I(\omega)} = 1 - Y_{\text{bound}} = 1-\alpha^2,
\end{equation}
which is shown as the black dotted line in Fig.~(\ref{fig:Continuum}).

\nocite{*}
\bibliography{Bibliography_abbreviated}% Produces the bibliography via BibTeX.

\end{document}